\documentstyle[preprint,aps]{revtex}

\begin{document}

\draft

%\onecolumn

\title
{Theory for Slightly Doped Antiferromagnetic Mott Insulators} 
\author{T.~K. Lee$^{1,2}$, Chang-Ming Ho$^{2}$, Naoto Nagaosa$^{3,4}$ 
and Wei-Cheng Lee$^{1}$}
\address{$^1$ Institute of Physics, Academia Sinica,
Nankang, Taipei, Taiwan 11529\\
$^2$ Physics Division, National Center for Theoretical Sciences,
P.O.Box 2-131,Hsinchu, Taiwan 300\\
$^3$ Department of Applied Physics, University of Tokyo, 7-3-1 Hongo,
Bunkyo-ku, Tokyo 113-8656, Japan\\
$^4$ Correlated Electron Research Center, AIST,
Tsukuba, Ibaraki 303-0046, Japan
}

\date{September, 2002. For the MOS2002 proceedings.}

\maketitle

\begin{abstract}
Trial wavefunctions, constructed explicitly from the unique 2-dimensional 
Mott insulating state with antiferromagnetic order, are proposed to describe 
the low-energy states of a Mott insulator slightly  doped  
with holes or electrons. With the state behaving like charged 
quasi-particles with well-defined momenta, a rigid band is observed. 
These states have much less pairing correlations than previously 
studied ones. Small Fermi patches obtained are consistent with
recent experiments on high $T_c$ cuprates doped lightly with holes or 
electrons. States showing the incoherent and spin-bag behaviors are 
also discussed. Using these wavefunctions, a number of results obtained 
by exact calculations are reproduced.  
\end{abstract}


With the continuous improvement of experimental technique and
sample making, it has recently become possible to study in detail
the phenomena in the lightly doped high $T_c$ cuprates. Indeed, many 
intriguing behaviors concerning the physics of doping the 
two-dimensional(2D) Mott insulator in this very undedoped regime 
are observed. Recent angle-resolved photoemission spectroscopy (ARPES) results unearth 
contrasting behaviors between lightly hole-doped $Ca_{2-z}Na_{z}CuO_2Cl_2$ 
($Na$-CCOC) \cite{Na-CCOC} and electron-doped $Nd_{2-x}Ce_xCuO_2$ (NCCO) 
high $T_c$ cuprates \cite{ncco-004}. 
Although ARPES on the undoped ({\em i.e.} $z$=$x$=0) insulating 
state shows an identical energy dispersion of a single hole created below the 
charge gap, results at a little higher dopings are demonstrated to be 
different: while a small hole patch is observed to be centered 
clearly at momentum ($\pi$/2,$\pi$/2) in 
the $Na$-CCOC at even $z$=0.1, small electron patches 
centered at ($\pi$,0) and (0,$\pi$) are observed for $x$=0.04 in NCCO.
Moreover, the in-plane transport in lightly doped systems shows high mobility of the charge carriers \cite{resistivity}. 
This and the recent high-resolution scanning tunneling microscopy/spectroscopy results indicate the signature 
of the {\em quasi-particle} behavior \cite{stm}. 

The single hole/electron behavior 
and its dispersion have been 
studied using various approaches on the $t$-$t'$-$t''$-$J$ model 
\cite{toh-mae94,xiang-wheatley,lee-shih,zk97,spin-bag,leung}. 
However, these studies on doping holes and electrons into the system 
only emphasize the {\em asymmetry} resulting from the different signs of 
$t'$ and $t''$ for the corresponding 
Hamiltonians. It is unclear whether the same physics is working for these 
two systems with {\em different} Hamiltonians. 
Should one try to construct a different theory
when electron-doped cuprates are considered?
Furthermore, do the models predict small Fermi surfaces and quasi-particles?

%There is another puzzle bothering many researchers in this area for many years.
%At half-filling, it has been shown by several 
%groups \cite{lee-feng,yokoyama-ogata}
%that the projected $d$-wave superconducting, or the resonating-valence-bond 
%(RVB), state with the antiferromagnetic long range order (AFLRO) to be an 
%excellent wavefunction (WF). In this state, the superconductivity (SC) is
%completely suppressed by the constraint of one particle per site, while
%the AFLRO survives. 

There are other issues concerning the model itself which have been bothersome. 
The studies applying the projected $d$-wave superconducting, or the 
resonating-valence-bond (RVB), state with the antiferromagnetic long range order (AFLRO)
also suggest that away from half-filling, 
the superconductivity (SC) revives and the ground state shows both SC and 
AFLRO\cite{lee-feng,himeda-ogata}. But so far most experiments do not 
support the coexistence. 
%This issue
%seriously undermines the validity of the $t$-$J$ type model. 
In addition, there has been not enough understanding on various 
properties found in the exact results of the $t$-$J$ type models on finite clusters \cite{toh-mae94,xiang-wheatley,zk97,spin-bag,leung,zk95,2-in-32} in terms of the many-particle wave functions (WF's).

In this paper, we discuss our recent proposal of a theory based on the variational 
approach to understand these 
issues \cite{lhn,llh}. Specific trial wavefunctions (TWF's) are constructed to describe 
both the low-energy states 
of the associated $t$-$t'$-$t''$-$J$ models with lightly doped holes and 
electrons. These  WF's are generalizations of the single-hole 
WF first written down by Lee and Shih \cite{lee-shih}.  
In contrast to other TWF's\cite{yokoyama-ogata}, ours are 
constructed solely from the optimized one at half-filling and include no 
hopping amplitudes $t'$ and $t''$.  
Yet, surprisingly, a number of properties including dispersion 
relations, momemtum distributions, spectral weight {\em etc.} are 
obtained correctly for both hole-doped and electron-doped 
systems. In the following, 
%we shall first present the mean-field theory and its TWF 
%at half-filling. 
%Then 
the {\em Lee-Shih WF} for one doped hole is used and
generalized for several holes as well as electron-doped systems.  
All numerical results reported below are from variational Monte 
Carlo (VMC) calculations for
an $8\times8$ lattice with periodic boundary conditions.

At half-filling, the system is described by the Heisenberg 
Hamiltonian  
\begin{equation}
{\cal H}_{J}=J\sum_{\langle i,j \rangle}({\bf S}_i\cdot{\bf S}_j-{1\over4} 
n_in_j),
\label{heisenberg}
\end{equation}
where $\langle i,j \rangle$ denotes nearest-neighbor ($n.n.$) sites.
Each site is occupied by only one single electron. As holes or electrons are 
doped into the system, we consider the Hamiltonian 
${\cal H}$=${\cal H}_{t-t'-t''}$+${\cal H}_{J}$ including $n.n.$ and longer 
range hoppings. Here 
${\cal H}_{t-t'-t''}$=$-t\sum_{\langle i,j\rangle \sigma}$ 
$\tilde{c}^{\dagger}_{i\sigma}\tilde{c}_{j\sigma}
-t'\sum_{\langle i,l\rangle \sigma}$ 
$\tilde{c}^{\dagger}
_{i\sigma}\tilde{c}_{l\sigma}-t''\sum_{\langle i,m\rangle \sigma}$ 
$\tilde{c}^{\dagger}
_{i\sigma}\tilde{c}_{m\sigma}$+H.c. with 
$\langle i,l \rangle$ and $\langle i,m \rangle$ representing the second $n.n.$ 
and third $n.n.$ site pairs. Note that $\tilde{c}_{i\sigma}$ in 
${\cal H}_{t-t'-t''}$ creates different kind of {\em holes} from 
single-electron-occupied sites at half-filling: 
{\em empty holes} (0{\it e}-hole) for hole doping 
and {\em two-electron-occupied holes} (2{\it e}-hole) for electron 
doping \cite{clarke}. 
Operator $\tilde{c}_{i\sigma}$ is actually equal to 
$c_{i\sigma}(1-n_{i,-\sigma})$ or $c_{i,-\sigma} n_{i\sigma}$ for hole or 
electron doped case, respectively. Therefore, despite the constraints, 
states in the two cases are in one-to-one correspondence after a local 
transformation $c_{i\sigma}\rightarrow c_{i,-\sigma}^{\dagger}$ is made. 
However, because of the Fermi statistics, 
the exchange of a single spin with a 2{\it e}-hole has an 
extra {\em minus} sign as compared to the 0{\it e}-hole. 
Hence, the only difference between the hole-doped and 
electron-doped $t$-$t'$-$t''$-$J$ model is $t'/t \rightarrow -t'/t$ and 
$t''/t \rightarrow-t''/t$
after we change the $c_{i\sigma}$ on B sublattice sites to $-c_{i\sigma}$ \cite{toh-mae94}. 
With all these, we then treat the hole and electron doped cases in the 
same manner with the Lee-Shih WF originally proposed only for a
 single hole. The VMC results presented in this paper are for
$J/t=0.3$, $t'/t=-(+)0.3$ and $t''/t=+(-)0.2$ in the hole(electron) 
doped case following the values usually used \cite{toh-mae94}.   

We shall apply the standard VMC method\cite{yokoyama-ogata} that enforces 
the local constraint exactly. Following the work by Lee and Shih 
\cite{lee-shih}, the TWF  for one doped hole with momentum 
$\bf q$ and $S_z$=$1/2$ is constructed to have $(N_s/2)-1$ singlet pairs of 
electrons and a single unpaired electron with momentum $\bf q$ 
and $S_z$=$1/2$,  
\begin{eqnarray}
|\Psi_1\rangle &=&P_d~c^{\dagger}_{{\bf q}\uparrow}
[{\sum_{\bf k} {}'
(A_{\bf k} a^{\dagger}_{{\bf k}\uparrow}a^{\dagger}_{{\bf -k}\downarrow}
 +B_{\bf k} b^{\dagger}_{{\bf k}\uparrow}b^{\dagger}_{{\bf -k}\downarrow})}
]^{(N_s/2)-1} 
|0\rangle . \nonumber 
\label{twf1}
\end{eqnarray}
The prime on the summation symbol indicates that the momentum $\bf
q$ is excluded from the sum if $\bf q$ is within the  sublattice 
Brillouin zone (SBZ),
otherwise, ${\bf q}-{\bf Q}$ is excluded. $N_s$ here is the total number of 
sites and {\bf Q}=$(\pi,\pi)$.

$|\Psi_{1}\rangle$ is explicitly constructed from the optimized
half-filled WF $|\Psi_0\rangle = P_d [{\sum_{\bf k} (A_{\bf k} 
a^{\dagger}_{{\bf k}\uparrow}a^{\dagger}_{{\bf -k}\downarrow}+B_{\bf k}
b^{\dagger}_{{\bf k}\uparrow}b^{\dagger}_{{\bf -k}\downarrow})}]^{N_s/2} 
|0\rangle $ and it does not contain any 
information about hoppings, $t'$, $t''$ and neither explicitly $t$, of the 
doped hole or electron. However,
the effect of $t$ is included in the RVB 
uniform bond 
$\chi$=$\langle \sum_{\sigma}c^{\dagger}_{i\sigma}c_{j\sigma}\rangle$
 which describes the large quantum fluctuation and spin singlet 
formation. There is also no need to introduce $t'$ and $t''$ in the TWF as they
are compatible with AFLRO. Here the coefficients 
$A_{\bf k}$
%=$(E_{\bf k}+\xi_{\bf k})/\Delta_{\bf k}$ 
and 
$B_{\bf k}$
%=$-(E_{\bf k}-\xi_{\bf k})/\Delta_{\bf k}$ 
are functions of $\xi_{\bf k}$ and $\Delta_{\bf k}$.
%$E_{\bf k}$=($\xi_{\bf k}^2+\Delta_{\bf k}^2$)$^{1/2}$. 
$\pm\xi_{\bf k}$=
$\pm(\epsilon_{\bf k}^2+(Jm_s)^2)^{1\over2}$ 
are energy dispersions for the two spin density wave (SDW) bands 
with 
$\epsilon_{\bf k}$=$-{3 \over 4}J\chi (cos{\rm k}_x+cos{\rm k}_y)$ and 
the staggered magnetization
$m_s$=$\langle S^z_A\rangle$=$-\langle S^z_B\rangle$, where the lattice 
is divided into A and B sublattices. 
$a_{{\bf k}\sigma}$ and $b_{{\bf k}\sigma}$ represent the operators 
of the lower and upper SDW bands, respectively,  and are 
related to the original electron operators $c_{{\bf k}\sigma}$ and 
$c_{{\bf k}+{\bf Q}\sigma}$ with  ${\bf Q}$=$(\pi,\pi)$ set for the 
commensurate SDW state. $\Delta_{\bf k}$=
${3\over 4}J\Delta d_{\bf k}$ with $d_{\bf k}$=$cos{\rm k}_x-cos{\rm k}_y$ 
here is for the $d$-wave RVB ($d$-RVB) order parameter.  
The projection operator $P_d$
enforces the constraint of no doubly occupied (or vacant) sites for cases 
with finite hole (or electron) doping . 
At half-filling, $N_s$ equals the total number of electrons. 
Notice that the sum in $|\Psi_0\rangle$ is taken over SBZ. 
There are two variational
parameters: $\Delta/\chi$ and $m_s/\chi$ in these WF's.

%%%%%%%%%%%%%%%%%% 1e-disp  %%%%%%%%%%%%%%%%%%

%\begin{figure}
%
%\centerline{\includegraphics*[width=80mm]{1e-disp.eps}}
%\centerline{\psfig{figure= ,width=8cm}}
%\caption{Energy dispersion of one electron in the $t$-$t'$-$t''$-$J$ 
%model on an $8\times8$ lattice. Black dots 
%are VMC results by using $|\Psi_1\rangle$. 
%The fitted dispersion $E_{1{\bf k}}-E_0$ are plotted as gray diamonds with 
%parameters $\chi=6.92$, $\Delta=2.71$, $m_s=18.84$, $E_0=7.43$, 
%$t_{eff}=0.06$, $t'_{eff}=-0.15$, $t''_{eff}=0.1$. Inset: patches in 
%one of BZ by filling the fitted dispersion in the main figure up to 
%$\sim$ 3\% doping.} 
%\label{1e-disp} \end{figure}

%%%%%%%%%%%%%%%%%% 1e-disp  %%%%%%%%%%%%%%%%%%

The energy dispersion obtained from  $|\Psi_{1}\rangle$ for one doped hole
has been shown by Lee and Shih\cite{lee-shih} to agree very well with
that of exact calculations, self-consistent Born approximations (SCBA),
and Green function Monte Carlo methods {\em etc.}. 
As for the case of having an extra up-spin electron with momentum ${\bf q}$ 
doped into the half-filled state, the energy dispersion can be calculated 
with this same WF $|\Psi_{1}\rangle$. The only 
difference is signs of  $t'/t$ and $t''/t$ are changed 
in the Hamiltonian.
  
The variational energies for one doped electron are shown as black dots 
in Fig.{\ref{1e-disp}}.  This result agrees well with that of Xiang and 
Wheatley\cite{xiang-wheatley} obtained by SCBA.
The optimal variational parameters are ($\Delta/\chi,m_{s}/\chi$)=(0.25,0.125).
The ground state  is at momentum 
${\bf q}=(\pi,0)$. The VMC results can be fitted simply by 
$E_{1{\bf k}}$=$E_{\bf k} -2t_{eff}(cos{\rm k}_x+cos{\rm k}_y)
-4t'_{eff}cos{\rm k}_xcos{\rm k}_y-2t''_{eff}
(cos(2{\rm k}_x)+cos(2{\rm k}_y))$
with parameters described in the caption of Fig.{\ref{1e-disp}}.  
The dispersion thus seems to be simply the combination of 
  the mean-field band at half-filling 
and the coherent hoppings \cite{lee-shih}.

To examine further 
the physical properties  of  $|\Psi_{1}\rangle$, we calculated the momentum 
distribution function $\langle n_{\sigma}^{h}({\bf k})\rangle$ for 
the ground state of a single hole with momentum  
${\bf q}=(\pi/2, \pi/2)$ and  $S_z$=$1/2$.
Results are shown in Fig.{\ref{nk1-eh}}(a) and (b). Note that the 
dips or pockets at $(\pi/2,\pi/2)$ and {\em anti-dips} at $(-\pi/2,-\pi/2)$ 
found by Leung \cite{leung} for the exact results of 32 sites are also 
clearly seen here.
It is quite amazing that $|\Psi_{1}\rangle$, including no 
$t'$ and $t''$, not only produces the correct energy dispersions 
for a single doped hole or electron it also provides a correct picture
about the momentum distribution.

%%%%%%%%%%%%%%%%%% nk1-eh %%%%%%%%%%%%%%%%%%

%\begin{figure}
%%
%%\centerline{\includegraphics*[width=80mm]{nk1-eh-c.eps}}
%\centerline{\psfig{figure=nk1-eh-c.eps,width=9cm}}
%\caption{Momentum distribution functions 
%$\langle n_{\sigma}^{h(e)}({\bf k})\rangle$ for a single hole, (a) and (b), 
%and electron, (c) and (d), in the 8$\times$8 $t$-$t'$-$t''$-$J$ model. 
%A scale is shown in between each set. 
%The darker area indicates smaller values of 
%$\langle n_{\sigma}^{h(e)}({\bf k})\rangle$. 
%} 
%\label{nk1-eh} \end{figure}

%%%%%%%%%%%%%%%%%% nk1-eh  %%%%%%%%%%%%%%%%%%

The momentum distribution functions $\langle n_{\sigma}^{e}({\bf k})\rangle$ 
for electron doped systems could be also calculated from  $|\Psi_{1}\rangle$
if we perform the transformation,  
 $c_{i\sigma}\rightarrow c_{i,-\sigma}^{\dagger}$,
 and $t$ is chosen to be positive.
In fact, it is easy to show that  $\langle n_{\sigma}^{e}({\bf k})\rangle$
$= 1 - \langle n_{-\sigma}^{h}({\bf Q}-{\bf k})\rangle$.
The results for the ground state 
of a single doped electron with momentum  ${\bf k}=(\pi,0)$ 
and spin $S_z$=$1/2$  are shown in Fig.{\ref{nk1-eh}}(c) and (d).
There are peaks at ${\bf k}=(\pi, 0)$ and an {\em anti-peak} at
${\bf k}=(0, \pi)$. 

Now we shall generalize the Lee-Shih WF 
$|\Psi_{1}\rangle$ to the case 
of two holes.  The simplest possible way is just to take out the
unpaired spin from  $|\Psi_{1}\rangle$ if we are interested in
the state with zero total momentum and $S_z$=$0$, which turns out to be
the lowest energy state. 
The TWF for  two holes with  momenta  ${\bf q}$ and  $-{\bf q}$ is 
\begin{eqnarray}
|\Psi_2\rangle &=& P_d[{\sum_{\bf k} {}' 
(A_{\bf k} a^{\dagger}_{{\bf k}\uparrow}
a^{\dagger}_{{\bf -k}\downarrow}
 +B_{\bf k} b^{\dagger}_{{\bf k}\uparrow}b^{\dagger}_{{\bf -k}\downarrow})}
]^{(N_s/2)-1} |0\rangle . \nonumber  
%\label{twf2}
\end{eqnarray}
Note that the momentum ${\bf q}$ is not included in the summation.  
It is most surprising to find that although $|\Psi_2\rangle$ has 
zero total momentum irrespective of ${\bf q}$,
its energy varies with the missing momentum or 
{\it the hole momentum} ${\bf q}$.
The dispersion 
%shown in Fig.{\ref{2eh-disp}}(a) is 
turns out to be very similar to
that of a single electron as shown in Fig.{\ref{1e-disp}} \cite{lhn}. 
The state with momentum ${\bf q}$=$(\pi,0)$ has the lowest energy for two 
electrons. The values of the two parameters $\Delta/\chi$ and $m_{s}/\chi$ 
are the same for $|\Psi_2\rangle$ and $|\Psi_1\rangle$.

%%%%%%%%%%%%%%%%%% 3eh-disp  %%%%%%%%%%%%%%%%%%

%\begin{figure}
%
%\centerline{\includegraphics*[width=60mm]{3eh-disp.eps}}
%\centerline{\psfig{figure=3eh-disp.eps,width=8cm}}
%\caption{Energy dispersions of (a) three doped electrons and 
%(b) three doped holes in
%the $t$-$t'$-$t''$-$J$ model. Filled circles are VMC results 
%for the momenta shown along the horizontal axis by using
%$|\Psi_3\rangle$. A total minus sign has been multiplied to
%the hole doped case. Open circles, with the momenta also 
%displayed, represent the lowest energy states.  
%Results here are obtained with parameters 
%%($\Delta/\chi,m_{s}/\chi$)=(0.25,0.125).
%shown in the figures.}
%\label{3eh-disp} 
%\end{figure}

%%%%%%%%%%%%%%%%%% 3eh-disp  %%%%%%%%%%%%%%%%%%

Using WF $|\Psi_2 \rangle$, the energy dispersion for two holes doped into the 
half-filled state 
%is shown in  Fig.{\ref{2eh-disp}}(b). Again its dispersion
again has an almost identical form as that of a single hole and the 
minimum is at ${\bf q}=(\pi/2,\pi/2)$ \cite{lee-shih}. 
The lowest energy obtained is $-26.438(3)t$ which
 is much lower than the variational energy,
$-25.72(1)t$, using the TWF applied by Himeda and 
Ogata \cite{himeda-ogata}.
Even if we include  $t'$ and $t''$, the variational energy 
$-25.763(7)t$, is still much higher than ours \cite{shih}.
%In the inset of  Fig.{\ref{2eh-disp}}
We also found that 
the hopping amplitudes for $n.n.$, second $n.n.$ and third $n.n.$ 
%are shown for one hole and two holes as a function of  ${\bf q}$. 
%It shows that the values 
of two holes are almost twice that of one hole.
The momentum distribution function for this state (not shown) has dips at
$(\pi/2,\pi/2)$ and $(-\pi/2,-\pi/2)$. This is in good agreement with
the exact result\cite{leung} for the $t$-$t'$-$t''$-$J$ model with 2 holes 
in 32 sites. 

%We have also examined the dispersion when
%the momenta of the two holes are chosen to be ${\bf q}$
%and  ${\bf p}\ne-{\bf q}$ with two unpaired spins. 
%The dispersion still works well.
%%Detail will be reported elsewhere.

%%%%%%%%%%%%%%%%%% 3eh-disp  %%%%%%%%%%%%%%%%%%

%\begin{figure}
%%\centerline{\includegraphics*[width=60mm]{3eh-disp.eps}}
%\centerline{\psfig{figure=3eh-disp.eps,width=8cm}}
%\caption{Energy dispersions of (a) three doped electrons and 
%(b) three doped holes in the $t$-$t'$-$t''$-$J$ model. Filled 
%circles are VMC results for the momenta shown along the 
%horizontal axis by using $|\Psi_3\rangle$. A total minus 
%sign has been multiplied to the hole doped case. Open circles, 
%with the momenta also displayed, represent the lowest energy 
%states. Results here are obtained with parameters shown in 
%the figures.}
%\label{3eh-disp} 
%\end{figure}

%%%%%%%%%%%%%%%%%% 3eh-disp  %%%%%%%%%%%%%%%%%%

It is then straightforward to write down the same type of TWF for three holes
with momenta ${\bf q}$, ${\bf q'}$ and $-{\bf q'}$:

\begin{eqnarray}
|\Psi_3\rangle &=& P_d~c^{\dagger}_{{\bf q}\uparrow}
[{\sum_{\bf k} {}' (A_{\bf k} a^{\dagger}_{{\bf k}\uparrow}a^{\dagger}_{{\bf -k}\downarrow}
 +B_{\bf k} b^{\dagger}_{{\bf k}\uparrow}b^{\dagger}_{{\bf -k}\downarrow})}]^{(N_s/2)-2} |0\rangle , 
\nonumber 
\label{twf3}
\end{eqnarray}
where ${\bf q'}$ and ${\bf q}$ are excluded from the summation. 
Just like the case with two holes or two electrons energy dispersions are 
proportional to the sum of the three single hole energies at momenta 
${\bf q}$, ${\bf q'}$ and $-{\bf q'}$. In Fig.{\ref{3eh-disp}}, the dispersions are plotted as functions of ${\bf q}$. For the electron doped ({\em i.e.} 
with 
three 2{\em e}-holes) case in Fig.{\ref{3eh-disp}}(a), the lowest energy 
state is at ${\bf q}=(3\pi/4,0)$ (shown as an open circle) 
within the SBZ after the first two electrons 
occupy ${\bf q'}=(\pi,0)$. With three doped holes (Fig.{\ref{3eh-disp}}(b)), after the two electrons at  
$(\pi/2,\pi/2)$ are removed the ground 
state is now at $(\pi/2,-\pi/2)$ (the open circle). As shown clearly in 
Fig.{\ref{3eh-disp}}, the dispersions 
follow nicely the single hole one.       

%In the Table, 
Values of the staggered magnetization 
$m$=${N_s}^{-1}{\sum}_{i}(-1)^{i}S_{i}^{z}$ are computed and compared for 
several 0{\it e}-hole and 2{\it e}-hole concentrations. 
%are compared. 
With the same variational parameters,  
it is found that the preference of $(\pi/2,\pi/2)$ for 0{\it e}-holes 
causes clearly larger disturbance of the AF order than for the electron 
doped case where 2{\it e}-holes with momentum 
$(\pi,0)$ shows much less influence on the AF order \cite{lhn}. This is consistent with
previous work \cite{toh-mae94}. It is also consistent with 
experimental results that
AF phase is more stable for electron doping than hole doping \cite{takagi}.

%%%%%%%%%%%%%%%%%%%%%%%%%%%%%%%%%%%%%%%%%%%%%%%%%%%%%%%%%%%%%%%%%%%%%

%\vspace{0.5cm}

%{\bf TABLE} Staggered magnetization $m$ for 1, 2 and 3 doped holes and electrons in an 8$\times$8 lattice.
%The parameters used here are ($\Delta/\chi,m_{s}/\chi$)=(0.25,0.125).
%\begin{center}
%\begin{tabular}{c | c c c c c} \hline
%\hline
%Doping number     & 0 & 1 & 2 & 3 \\
%\hline
%Hole doped  &0.365(1)&0.353(1)& 0.329(1)&0.285(1) \\
%\hline
%Electron doped  &0.365(1) & 0.372(1) & 0.348(1) & 0.332(4) \\
%\hline
%\end{tabular}
%\end{center}
%\vspace{0.5cm}

%%%%%%%%%%%%%%%%%%%%%%%%%%%%%%%%%%%%%%%%%%%%%%%%%%%%%%%%%%%%%%%%%%%%%%%        

So far, based on the  $t$-$t'$-$t''$-$J$ model we have proposed
 a TWF to describe the low energy states 
of slightly doped  antiferromagnetic Mott insulators.
Exactly the same TWF's are proposed to account for the behavior of
both hole doping and electron doping, after we employed the 
particle-hole transformation. Different energy dispersions for these two cases
are due to the different signs of $t'/t$ and $t''/t$ which is a direct 
consequence of the constraint that electron doped system has 
2{\it e}-holes while hole doped system only has 
0{\it e}-holes. Rigid band and {\em quasi-particle}
 behavior are demonstrated for both cases. 
The theory provides an explanation of recent ARPES results. 
In lightly hole-doped cuprates, small Fermi pocket is around 
$(\pi/2,\pi/2)$. In electron-doped cuprates, the patch is around 
$(\pi,0)$ as shown in the inset of Fig.{\ref{1e-disp}} with doping at 
about {3\%}. 

%%%%%%%%%%%%%%%%%% 4h-hh  %%%%%%%%%%%%%%%%%%
%\begin{figure}
%
%\centerline{\includegraphics*[width=80mm]{4h-hh.eps}}
%\centerline{\psfig{figure=4h-hh.eps,width=8cm}}
%\caption{Hole-hole correlation functions with 4 doped holes in the 
%8$\times$8 lattice. The result obtained using our WF, {\em i.e.} same form
%as $|\Psi_2 \rangle$ with momenta ($\pi$/2,$\pi$/2) and ($\pi$/2,$-\pi$/2)
%excluded in the sum, is compared with that using the Himeda-Ogata
%WF \cite{himeda-ogata}. Results here are obtained with optimized
%parameters ($\Delta/\chi,m_{s}/\chi$)=(0.25,0.125) and,
%for Himeda-Ogata one, also $\mu$=-0.025.}
%\label{4h-hh} 
%\end{figure}
%%%%%%%%%%%%%%%%%% 4h-hh  %%%%%%%%%%%%%%%%%%

Another important property of our WF's is that
holes are essentially independent of each other as they obey the same energy
dispersions (with very little renormalization of parameters). Exactly because
this quasi-particle like property is unchanged after doping,
our WF has AFLRO but very little
superconducting pairing correlations.
The presence of superconducting state certainly will change
the excitation spectra. In particular, the $d$-wave SC,
which coexists with AFLRO in some of
the previous variational studies, should have low energy excitations
along the nodes. This is certainly not seen in our TWF's.
In addition, the holes are not attractive to each other
in our WF's.
In Fig.{\ref{4h-hh}} the hole-hole correlation function for our
TWF and the WF used by authors in \cite{himeda-ogata} are compared
for 4 holes in an 8$\times$8 lattice.
The lack of attraction between
holes is consistent with Leung's low energy states obtained
exactly for two holes in 32 sites\cite{leung}.
Long range $d$-wave pairing correlation defined \cite{yokoyama-ogata}
  for our TWF and that in Himeda-Ogata one are, on average, about $0.002$, and $0.018$, respectively.
 Thus the $d$-RVB pairing for spins assumed by our WF's
are not in any way implying the pairing of charges.

%%%%%%%%%%%%%%%%%% qp-sb  %%%%%%%%%%%%%%%%%%
%\begin{figure}
%
%\centerline{\includegraphics*[width=80mm]{qp-sb.eps}}
%\caption{Energy dispersions of single doped hole (0{\em e} hole) in the
%8$\times$8 lattice using TWF's $|\Psi_1^{'} \rangle$ and $|\Psi_1 \rangle$.
%A total minus sign has been mutiplied here.
%The results obtained using WF $|\Psi_1^{'}\rangle$ (...)  are with fixed ${\bf \
%q}_{h}=$($\pi$/2,$\pi$/2). The legend along the horizontal axis then represnts \
%${\bf q}_{s}$, momenta of the unpaired spin. For $|\Psi_1 \rangle$ (...),  ${\bf q}_{s}=
%{\bf q}_{h}={\bf q}$. Lines in the figure serve as only guide to the eyes. Possible continuum states are represented by the shaded region. Results here are o\
%btained with parameters ($\Delta/\chi,m_{s}/\chi$)=(0.25,0.125).}
%\label{qp-sb}
%\end{figure}
%%%%%%%%%%%%%%%%%% qp-sb  %%%%%%%%%%%%%%%%%%

%%%%%%%%%%%%%%%%%% 4h-hh  %%%%%%%%%%%%%%%%%%^M

%\begin{figure}

%\centerline{\includegraphics*[width=80mm]{4h-hh.eps}}
%\centerline{\psfig{figure=4h-hh.eps,width=8cm}}
%\caption{Hole-hole correlation functions with 4 doped holes 
%in the 8$\times$8 lattice. The result obtained using our WF, 
%{\em i.e.} same form as $|\Psi_2 \rangle$ with momenta 
%($\pi$/2,$\pi$/2) and ($\pi$/2,$-\pi$/2) excluded in the sum, 
%is compared with that using the Himeda-Ogata WF 
%\cite{himeda-ogata}
%. Results here are obtained with optimized
%parameters ($\Delta / \chi,m_{s} / \chi$)=(0.25,0.125) and, 
%for Himeda-Ogata one, also $\mu$=-0.025.}
%\label{4h-hh} 
%\end{figure}
%%%%%%%%%%%%%%%%%% 4h-hh  %%%%%%%%%%%%%%%%%%^M

Although our TWF's or the Lee-Shih WF's has reproduced many 
numerical results obtained by exact diagonalization, 
SCBA, {\em etc.} for one or two holes or electrons, 
%it provides several new results. 
there exists, however, some inconsistancy in the detail of the 
comparison. Namely, at some momenta ${\bf k}$'s the spectral weights 
$Z_{\bf k}=|\langle \Psi_{\bf k} | c_{{\bf k}\sigma} | \Psi_{0} 
\rangle|^{2} / \langle \Psi_{0} | c^{\dagger}_{{\bf k}\sigma}
c_{{\bf k}\sigma} | \Psi_{0} \rangle$ computed for the one doped 
hole case using $|\Psi_{\bf k} \rangle = |\Psi_{1} \rangle$ are 
much larger than the exact results \cite{zk97}. This may indicate that 
the quasi-particle states do not exist everywhere in the Brillouin zone. In 
fact, the ARPES on the undoped ($z=0$) $Na$-CCOC also show well-defined 
peaks only locally in the high-symmetry directions \cite{Na-CCOC}. 

We have constructed TWF's with which small spectral weights are realized 
at particular ${\bf k}$ points \cite{llh}. For the single hole doped case, 
state described by WF   
\begin{eqnarray}
|\Psi_1^{'}\rangle &=&P_d~c^{\dagger}_{{\bf q}_{s}\uparrow}
[{\sum_{[{\bf k} \neq {\bf q}_{h}]} {}'
(A_{\bf k} a^{\dagger}_{{\bf k}\uparrow}a^{\dagger}_{{\bf -k}\downarrow}
 +B_{\bf k} b^{\dagger}_{{\bf k}\uparrow}b^{\dagger}_{{\bf -k}\downarrow})}
]^{(N_s/2)-1}
|0\rangle  \nonumber
\label{twf1'}
\end{eqnarray}
with the hole momentum ${\bf q}_{h}=(\pi/2,\pi/2)$ has also one singlet bond 
less than the half-filled case, but the unpaired spin momentum 
${\bf q}_{s}$ is chosen here not to be the same as ${\bf q}_{h}$. 
Note that only ${\bf q}_{h}$ is excluded within the sum 
for $|\Psi_1^{'}\rangle$.
With the same parameters $\Delta/\chi$ and $m_{s}/\chi$, it is 
found that $|\Psi_1^{'}\rangle$ can have {\em lower} variational energy 
than that of the quasi-particle state $|\Psi_1 \rangle$ at 
some ${\bf q}_{s}$'s. Apparently, many states could be constructed
with same ${\bf q}_{s}$ but different ${\bf q}_{h}$. These states 
constitutes the continuum at {\em higher} energy.  
      
Substituting WF $|\Psi_1^{'} \rangle$ and the quasi-particle 
$|\Psi_{1} \rangle$ into $|\Psi_{\bf k} \rangle$ and computing 
$Z_{\bf k}$ suffices to reproduce 
the variation of the spectral weights obtained in the exact 
results \cite{zk97}. States described by  $|\Psi_1^{'} \rangle$ 
have smaller $Z_{\bf k}$'s and are thus with a 
strong {\em incoherent} character. 
We also found that 
more quasi-particle states
%, {\em i.e.} with large-weighted $Z_{\bf k}$'s, 
have lower energy below the continuum 
in the pure $t$-$J$ model with $t'=t''=0$ \cite{zk95} than in 
the $t$-$t'$-$t''$-$J$ model. This is consistent 
with what has been known in the exact results \cite{zk97}. 

To understand the incoherent states more, the hole-spin correlations for 
$|\Psi_1^{'} \rangle$ with different ${\bf q}_{s}$'s are examined \cite{llh}. 
The spin configurations around the hole are quite different 
from that of the 
quasi-particle states. Spin moments around the hole in the 
quasi-particle state $|\Psi_1 \rangle$ can have values larger than the 
average at that momentum, thus the unpaired spin seems to be bound to
the hole. However, in the incoherent state the staggered magnetization
is suppressed around the hole, this is like a spin-bag state\cite{sch}.
The unpaired spin is no longer associated with the hole and this is
now a spin-charge separated state.
The amazing difference between these two wave functions could
be understood quite easily. 
When we apply  the spin flipping operator,
$S^{\dagger (-)}({\bf k}^{'};{\bf q}^{'} ={\bf q}[={\bf q}_{h}])=\sum_{{\bf q}^{'}} c^{\dagger}_{{\bf q}^{'} + {\bf k}^{'} \uparrow(\downarrow)} c_{{\bf q}^{'} \downarrow(\uparrow)} |_{{\bf q}^{'} ={\bf q}}$ 
with ${\bf q}^{'} + {\bf k}^{'}={\bf q}_{s}$, 
to the quasi-particle state, 
 the unpaired spin ${\bf q}$ is excited to a new momentum ${\bf q}_{s}$
 in the state $|\Psi_1^{'} \rangle$. Thus the unpaired spin is no longer
bound to the hole. 
 Furthermore, the spin-spin correlations 
across the hole with ${\bf q}_s=(\pi,0)$ show the {\em anti-phase} domain, 
{\em i.e.} AF correlation at the same sublattice. These features 
reproduce what have been obtained in the exact calculations \cite{spin-bag}.
 
In summary, we have discussed new TWF's describing the low-energy states 
of the $t$-$t'$-$t''$-$J$ model at lightly doping. These states reproduce
various exact numerical results and, also, show consistent behaviors with 
what have been observed in the experiments.

\smallskip

\noindent
We are grateful to Drs. A. Damascelli, F. Ronning and Prof. Z.-X. Shen for 
sharing their ARPES data prior to publication, and Profs. C.-T. Shih and P.~W. Leung for some supporting results and invaluable 
discussions. TKL is supported by the grant NSC89-2112-M-001-103 (R.O.C.) and 
NN is supported by Priority Areas Grants and Grant-in-Aid for COE research 
from the Ministry of Education, Culture, Sports, Science and Technology 
of Japan.

%\vspace{-0.7cm}

%\end{thebibliography}

\newpage

%%%%%%%%%%%%%%%%%% 1e-disp  %%%%%%%%%%%%%%%%%%^M

\begin{figure}
%^M
%\centerline{\includegraphics*[width=80mm]{1e-disp.eps}}
%\centerline{\psfig{figure= ,width=8cm}}
\caption{Energy dispersion of one electron in the $t$-$t'$-$t''$-$J$
model on an $8\times8$ lattice. Black dots
are VMC results by using $|\Psi_1\rangle$.
The fitted dispersion $E_{1{\bf k}}-E_0$ are plotted as gray diamonds with
parameters $\chi=6.92$, $\Delta=2.71$, $m_s=18.84$, $E_0=7.43$,
$t_{eff}=0.06$, $t'_{eff}=-0.15$, $t''_{eff}=0.1$. Inset: patches in
one of BZ by filling the fitted dispersion in the main figure up to
$\sim$ 3\% doping.}
\label{1e-disp} \end{figure}

%%%%%%%%%%%%%%%%%% 1e-disp  %%%%%%%%%%%%%%%%%%

%%%%%%%%%%%%%%%%%% nk1-eh %%%%%%%%%%%%%%%%%%

\begin{figure}

%\centerline{\includegraphics*[width=80mm]{nk1-eh-c.eps}}
%\centerline{\psfig{figure=nk1-eh-c.eps,width=9cm}}
\caption{Momentum distribution functions 
$\langle n_{\sigma}^{h(e)}({\bf k})\rangle$ for a single hole, 
(a) and (b), and electron, (c) and (d), in the 
8$\times$8 $t$-$t'$-$t''$-$J$ model. A scale is shown in between 
each set. The darker area indicates smaller values of 
$\langle n_{\sigma}^{h(e)}({\bf k})\rangle$.} 
\label{nk1-eh} 
\end{figure}

%%%%%%%%%%%%%%%%%% nk1-eh  %%%%%%%%%%%%%%%%%%^M

%%%%%%%%%%%%%%%%%% 3eh-disp  %%%%%%%%%%%%%%%%%%^M

\begin{figure}
%\centerline{\includegraphics*[width=60mm]{3eh-disp.eps}}
%\centerline{\psfig{figure=3eh-disp.eps,width=8cm}}
\caption{Energy dispersions of (a) three doped electrons and
(b) three doped holes in the $t$-$t'$-$t''$-$J$ model. Filled
circles are VMC results for the momenta shown along the
horizontal axis by using $|\Psi_3\rangle$. A total minus
sign has been multiplied to the hole doped case. Open circles,
with the momenta also displayed, represent the lowest energy
states. Results here are obtained with parameters shown in
the figures.}
\label{3eh-disp}
\end{figure}

%%%%%%%%%%%%%%%%%% 3eh-disp  %%%%%%%%%%%%%%%%%%^M

%%%%%%%%%%%%%%%%%% 4h-hh  %%%%%%%%%%%%%%%%%%^M

\begin{figure}

%\centerline{\includegraphics*[width=80mm]{4h-hh.eps}}
%\centerline{\psfig{figure=4h-hh.eps,width=8cm}}
\caption{Hole-hole correlation functions with 4 doped holes
in the 8$\times$8 lattice. The result obtained using our WF,
{\em i.e.} same form as $|\Psi_2 \rangle$ with momenta
($\pi$/2,$\pi$/2) and ($\pi$/2,$-\pi$/2) excluded in the sum,
is compared with that using the Himeda-Ogata WF
%\cite{himeda-ogata}
. Results here are obtained with optimized
parameters ($\Delta / \chi,m_{s} / \chi$)=(0.25,0.125) and,
for Himeda-Ogata one, also $\mu$=-0.025.}
%\cite{himeda-ogata}
\label{4h-hh}
\end{figure}
%%%%%%%%%%%%%%%%%% 4h-hh  %%%%%%%%%%%%%%%%%%^M


\begin{references}
\bibitem{shih} Actually, the lowest energy for Himeda-Ogata type TWF 
is obtained when its parameters are chosen to replicate our TWF. 
Details are to be reported elsewhere.
\end{references}
\end{document}